\documentclass[reprint,aps,prl,twocolumn,superscriptaddress,nobibnotes,floatfix]{revtex4-1}%

\usepackage{amsmath}
\usepackage{amssymb}
\usepackage{natbib}
\usepackage{graphicx}
\usepackage{xcolor}
\usepackage[unicode=true, breaklinks=false, pdfborder={0 0 1}, backref=false, colorlinks=true, linkcolor=blue, urlcolor=blue, citecolor=blue]{hyperref}

\setcitestyle{numbers,square}
\begin{document}


\title{Nematicity Arising from a Chiral Superconducting Ground State in Magic-Angle Twisted Bilayer Graphene under In-Plane Magnetic Fields}

\author{Tao Yu}
\affiliation{Max Planck Institute for the Structure and Dynamics of Matter, Luruper Chaussee 149, 22761 Hamburg, Germany}
\author{Dante M. Kennes}
\affiliation{Institut f\"ur Theorie der Statistischen Physik, RWTH Aachen University and JARA-Fundamentals of Future Information Technology, 52056 Aachen, Germany}
\affiliation{Max Planck Institute for the Structure and Dynamics of Matter, Luruper Chaussee 149, 22761 Hamburg, Germany}
\author{Angel Rubio}
\affiliation{Max Planck Institute for the Structure and Dynamics of Matter, Luruper Chaussee 149, 22761 Hamburg, Germany}
\affiliation{Center for Computational Quantum Physics (CCQ), The Flatiron Institute,
	162 Fifth Avenue, New York, New York 10010, USA}
	\affiliation{Nano-Bio Spectroscopy Group, Departamento de Física de Materiales,
	Universidad del País Vasco, 20018 San Sebastian, Spain}
\author{Michael A. Sentef}
\affiliation{Max Planck Institute for the Structure and Dynamics of Matter, Luruper Chaussee 149, 22761 Hamburg, Germany}

\date{\today}

\begin{abstract}
	Recent measurements of the resistivity in magic-angle twisted bilayer graphene near the superconducting transition temperature show two-fold anisotropy, or nematicity, when changing the direction of an in-plane magnetic field [Cao \textit{et al.}, Science \textbf{372}, 264 (2021)]. This was interpreted as strong evidence for exotic nematic superconductivity instead of the widely proposed chiral superconductivity. Counter-intuitively, we demonstrate that in two-dimensional chiral superconductors the in-plane magnetic field can hybridize the two chiral superconducting order parameters to induce a phase that shows nematicity in the transport response. Its paraconductivity is modulated as $\cos(2\theta_{\bf B})$, with $\theta_{\bf B}$ being the direction of the in-plane magnetic field, consistent with experiment in twisted bilayer graphene. We therefore suggest that the nematic response reported by Cao \textit{et al.} does not rule out a chiral superconducting ground state.
\end{abstract}
\maketitle

\textit{Introduction}.---The pairing symmetries are fundamental properties of the superconducting state and yield robust insights even irrespective of the details of the underlying microscopic pairing mechanisms \cite{RMP,pairing_cuprate,pairing_triplet,SC_symmetry}. The recently discovered superconducting phase close to the correlated insulating phase in magic-angle twisted bilayer graphene (MATBG) \cite{NS1,NS2,NS3,NS4} has spurred tremendous research activities. However, the pairing symmetry of the superconducting state has not been identified experimentally. Pairing mechanisms based on phonons \cite{phonon1,phonon2,phonon3,phonon4,phonon5,phonon6} or pure Coulomb interaction \cite{Coulomb1,Coulomb2,Coulomb3,Coulomb4,Coulomb5,Coulomb6,Coulomb7,Coulomb8,Coulomb9,Coulomb10,Coulomb11,Coulomb12,Coulomb13,Coulomb14,Coulomb15,Coulomb16,Martin,Senthil,Coulomb17} have been proposed, among which the pure electronic origins often favor the chiral ($d\pm id$)-wave superconductivity with promising applications in topological quantum computing.  Chiral ($d\pm id$)-wave superconductivity retains the rotational symmetry but breaks the time-reversal one, by mechanism similar to the one previously studied in heavily doped single layer graphene \cite{graphene_NP,Honerkamp}. The existing evidence for chiral superconductivity in other materials, such as UPt$_3$ \cite{UPt1,UPt2,UPt3} and UTe$_2$ \cite{UTe}, has so far not been reported in MATBG.

Recent transport measurements in MATBG  provided key features of the pairing symmetry of the superconducting state by revealing a two-fold anisotropy or nematicity in the resistivity around the superconducting transition temperature $T_c$ when changing the direction of a relatively-strong ($\gtrsim 0.5$~T) in-plane magnetic field \cite{Pablo_exp}. The transport response is still isotropic when the magnetic field is smaller. At first glance it appears that chiral superconductivity should be ruled out since it respects the three-fold rotation symmetry of the lattice. Nematic superconductivity---an exotic phase that breaks the lattice rotational symmetry but respects the translational one---may be favored, which was phenomenologically proposed to be a complicated coexisting phase \cite{nematic_superconductivity} or intrinsic correlated phase \cite{nematic_correlated1}. Nematic fluctuation in the correlated insulating phases was indeed observed in MATBG by scanning tunneling microscopy (STM) \cite{STM_nematicity1,STM_nematicity2,STM_nematicity3}, also in twisted double bilayer graphene \cite{tDBG_nematicity}. But it is not very clear whether the insulating correlated phase is directly related to, or purely competitive with, the superconducting one, in that superconductivity can survive even when the insulating state is completely suppressed \cite{screening}. Less attention was, however, paid to the possible role of the in-plane magnetic field for superconductivity. As one marked exception, it was proposed to provide a vector potential to induce the $Z_2$ symmetry-breaking phase transition in Sr$_2$RuO$_4$ films of $(p\pm ip)$-wave chiral superconductors \cite{Silaev}. However, this was not observed since Sr$_2$RuO$_4$ may not be a $p$-wave superconductor, as suggested by recent investigations \cite{rule_out1,rule_out2,rule_out3}.

 \begin{figure}[t]
 	\begin{center}
 		{\includegraphics[width=4.7cm,trim=0.9cm 0cm 0.6cm 0cm]{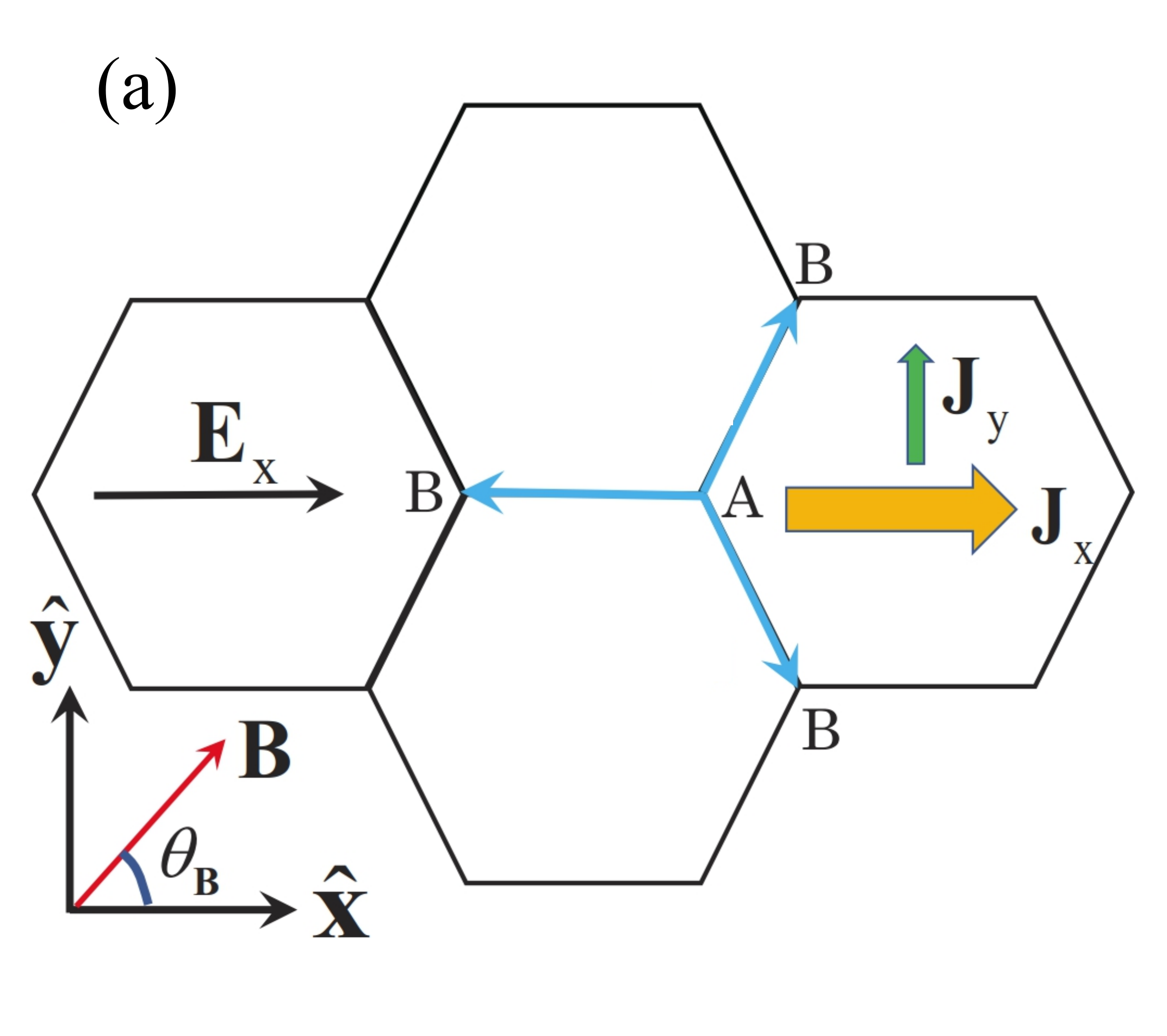}}
 		{\includegraphics[width=3.85cm,trim=0.45cm 0cm 0.5cm 0cm]{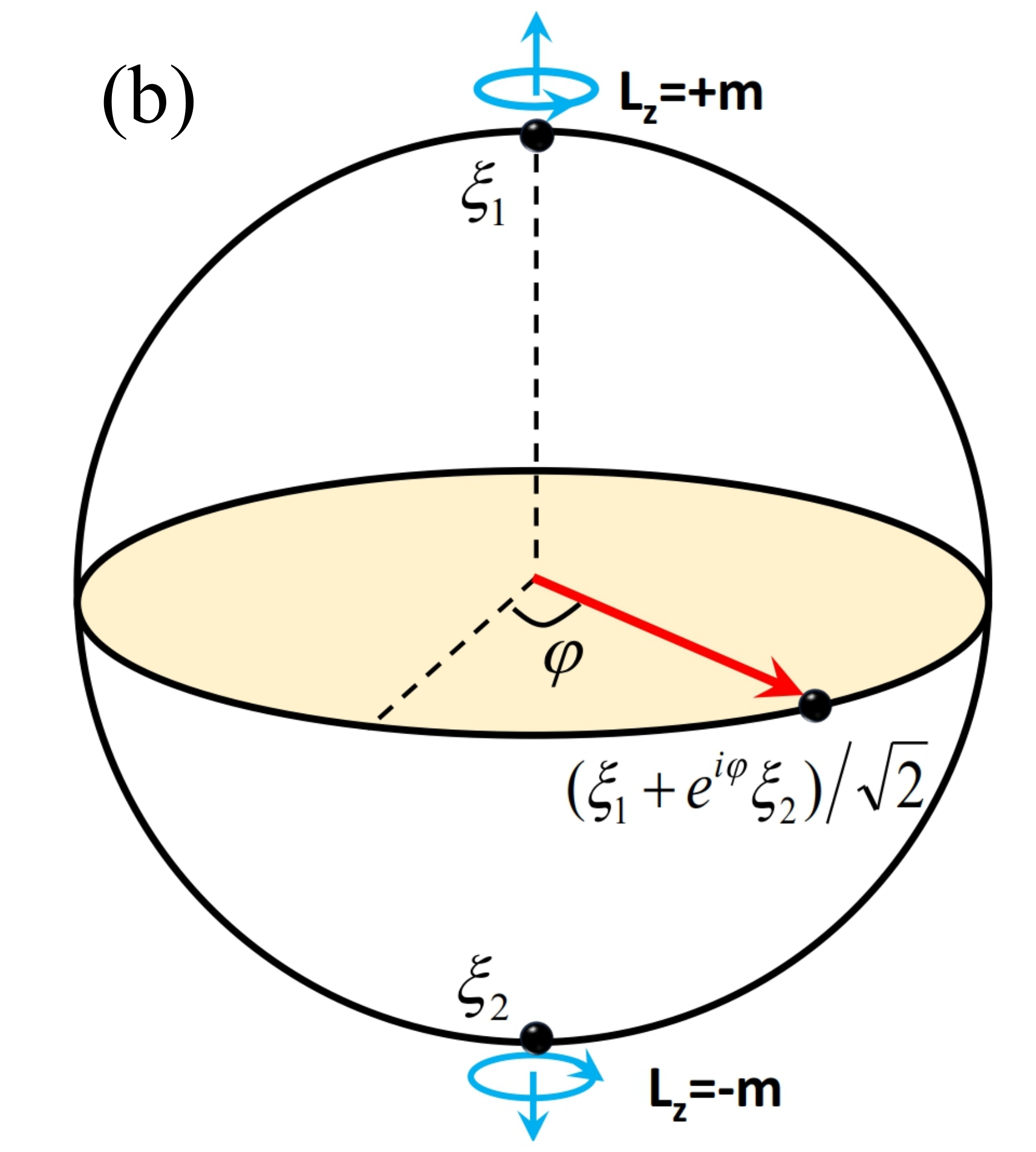}}
 		\caption{Nematic and Hall responses of the driven state by an in-plane magnetic field $\textbf{B}$ [(a)] via the hybridization of two degenerate chiral states on the Bloch sphere [(b)]. In (a), both the longitudinal current $\textbf{J}_x$ and Hall current $\textbf{J}_y$ under the electric field $\textbf{E}_x$ show two-fold anisotropy with respect to the magnetic-field direction $\theta_{{\bf B}}$.
 	    In (b), the north and south poles represent two  chiral states $\xi_{1,2}$ with positive and negative angular momentum $\pm m$ and the other points represent their superpositions. The in-plane magnetic field can drive the system to be a state $(\xi_1+\exp(i\varphi)\xi_2)/\sqrt{2}$ with equal contributions from the two chiral states, represented by points at the equator and modulated by $\varphi=2\theta_{\bf B}$.}  
 		\label{fig:Bloch}
 	\end{center}
 \end{figure}

In this Letter, we formulate the phase transition of a chiral ($d\pm id$)-wave superconductor driven by a critical in-plane magnetic field in a prototype honeycomb lattice of MATBG and demonstrate that the new phase is nematic with two-fold anisotropy in the transport response. We predict an emerging  Hall effect \cite{AHE_RMP} in the paraconductivity of the driven nematic phase, also with a two-fold anisotropy with respect to the direction of the in-plane magnetic field [Fig.~\ref{fig:Bloch}(a)]. In detail, the two degenerate chiral states, represented by $\xi_1$ and $\xi_2$ at the north and south poles of the Bloch sphere depicted in Fig.~\ref{fig:Bloch}(b), are coupled via the vector potential of the magnetic field through angular-momentum conservation. When the magnetic field $B$ is larger than a critical one $B_c$, the two chiral states are hybridized with equal contributions of the form $[\xi_1+\exp({2i\theta_{\bf B}})\xi_2]/\sqrt{2}$, as depicted by points at the equator of the Bloch sphere. The coefficient of this superposition is modulated by the direction of the magnetic field denoted by angle $\theta_{\bf B}$. Near the superconducting transition temperature $T_c$ the critical field driving the transition becomes arbitrarily small $B_c\to 0$ as $T\to T_c$ (e.g., $B_c\sim 0.6$~T when $T\sim 0.9 T_c$).  Although the chiral states $\xi_1$ and $\xi_2$ are both isotropic, the driven state given by their superposition is nematic with an anisotropy axis modulated by $\cos(2\theta_{\bf B})$, i.e., showing two-fold nematic response with respect to the applied field. The consistence with experimental measurements \cite{Pablo_exp} indicates that chiral superconductivity might be not ruled out in MATBG by these experimental findings.
 We propose that magnetoelectric transport measurements are useful tools also for other possible $(p\pm ip)$-wave chiral superconductors, such as UPt$_3$ \cite{UPt1,UPt2,UPt3} and UTe$_2$ \cite{UTe} thin films, to engineer and identify the pairing symmetry.

\textit{Symmetry analysis for magnetic-field driven nematicity.}---Although the microscopic mechanism of superconductivity in MATBG may be sensitive to the detailed structure of the flat bands and many-body interaction \cite{phonon1,phonon2,phonon3,phonon4,phonon5,phonon6,Coulomb1,Coulomb2,Coulomb3,Coulomb4,Coulomb5,Coulomb6,Coulomb7,Coulomb8,Coulomb9,Coulomb10,Coulomb11,Coulomb12,Coulomb13,Coulomb14,Coulomb15,Coulomb16,Martin,Senthil,Coulomb17}, the Ginzburg-Landau (GL) phenomenology is largely independent of these details by relying solely on the system's symmetry  \cite{RMP,SC_symmetry}. For MATBG, the (emergent) $D_6$ symmetry, which contains a six-fold rotation around the normal and two-fold rotations around in-plane axes, is successfully adopted to describe the band structure in the continuum model and tight-binding model with lattice relaxation \cite{continuum,Yuan_Fu2,Senthil,D6_relaxation}. For our purpose, we shall focus on the $d$-wave superconductivity, which is allowed in $D_6$ group by the two-dimensional irreducible representation with basis $\xi_{1,2}$ in form of $(d_{x^2-y^2}\pm id_{xy})$-waves,
with which one can represent the condensate Bose field via  introducing the superconducting order parameters $\psi_{1,2}({\bf r},t)$ under the basis $\xi_{1,2}$  \cite{RMP,SC_symmetry,Honerkamp}: 
\begin{align}
	\Phi({\bf r},t)=\psi_1({\bf r},t)\xi_1+\psi_2({\bf r},t)\xi_2.
	\label{decomposition}
\end{align}
The two basis functions  $\xi_{1,2}$ span a two-dimensional space, depicted as a Bloch sphere in Fig.~\ref{fig:Bloch}(b), with the $d_{x^2-y^2}\pm id_{xy}$ states being the north and south poles.  To construct the GL Lagrangian, we build the symmetry-allowed quadratic, quartic, and gradient terms of order parameters (refer to Supplemental Material for details \cite{supplement}). We note that symmetry analysis by the $D_3$, $D_6$ and $D_{6h}$ groups gives the same Lagrangian \cite{supplement,RMP,polynomial,Honerkamp}. The applied \textit{static} and \textit{uniform} in-plane magnetic field ${\bf B}$ induces an effective in-plane vector potential ${\bf A}_{x,y}$ via an average over the thickness $d$ of MATBG \cite{Silaev}, i.e.,
\begin{align}
	{\bf A}_{x,y}\rightarrow\sqrt{\langle A_{x,y}^2\rangle}=|B_{y,x}|d/\sqrt{6},
\end{align} 
where $\langle\cdots\rangle=\int_0^d dz(\cdots)$ denotes the spatial average over the sample thickness and ${\bf A}={\bf z}\times {\bf B}$ with the surface normal of the film along the $\hat{\bf z}$-direction. Including this vector potential, the GL Lagrangian density follows as 
\begin{align}
	\nonumber
	{\cal L}_{\rm eff}({\bf r})&=\alpha\sum_{\mu=1,2}|\psi_{\mu}({\bf r})|^2\\
	\nonumber
	&+\beta_{1}\left(D_{+}\psi_1\tilde{D}_{-}\psi_1^*+D_{-}\psi_2\tilde{D}_+\psi_2^*\right)\\
	\nonumber
	&+\beta_{2}\left(D_{-}\psi_1\tilde{D}_{+}\psi_1^*+D_{+}\psi_2\tilde{D}_-\psi_2^*\right)\\
	&+\gamma \left(D_+\psi_1
	\tilde{D}_+\psi_2^*+D_-\psi_2\tilde{D}_-\psi^*_1\right)
	\nonumber\\
	\nonumber
	&+\lambda_1\left(|\psi_1({\bf r})|^2+|\psi_2({\bf r})|^2\right)^2\\
	&+\lambda_2\left(|\psi_1({\bf r})|^2-|\psi_2({\bf r})|^2\right)^2,
	\label{free_energy}
\end{align}
where $\{\alpha,\beta_{1,2},\gamma,\lambda_{1,2}\}$ are real GL parameters, $D_{\pm}=\partial_{\pm}-(2e/i\hbar){\bf A}_{\pm}$, and $\tilde{D}_{\pm}=\partial_{\pm}+(2e/i\hbar){\bf A}_{\pm}$,  with $\partial_{\pm}\equiv\partial_x\pm i\partial_y$ and ${\bf A}_{\pm}\equiv{\bf A}_x\pm i{\bf A}_y$. 
The Lagrangian is invariant under the gauge transformation
\begin{align}
	\nonumber
	\psi_{1,2}({\bf r})\rightarrow \psi_{1,2}({\bf r})e^{i\Lambda({\bf r})},\quad {\bf A}({\bf r})\rightarrow {\bf A}({\bf r})-({\hbar}/{2e})\nabla\Lambda({\bf r}),
\end{align}
where $\Lambda({\bf r})$ is an arbitrary function. A constant vector potential is equivalent to a uniform supercurrent, to which case the conclusion drawn by magnetic field can be also applied. Without the magnetic field, the mass term $\alpha$ and stiffness $\beta_{1,2}$ are isotropic for either $\psi_1$ or $\psi_2$, viz., they are isotropic phases without nematicity. The two chiral order parameters are coupled via the orbital effect of a  magnetic field.

Under the application of an in-plane magnetic field, the ground state can be changed, which is found by minimizing the free energy Eq.~(\ref{free_energy}), yielding
\begin{align}
	\nonumber
	&\left(\begin{matrix}
		\alpha+\beta\left({2 e}/{\hbar}\right)^2 {\bf A}^2&-\gamma \left({2 e}/{\hbar }\right)^2{\bf A}^2e^{-i\varphi}\\
		-\gamma\left({2 e}/{\hbar }\right)^2{\bf A}^2e^{i\varphi }&\alpha+\beta\left({2 e}/{\hbar }\right)^2{\bf A}^2
	\end{matrix}\right)\left(\begin{matrix}
		\psi_1\\
		\psi_2
	\end{matrix}\right)\\
	&+\left(\begin{matrix}
		\Omega_+[\psi_1,\psi_2]&0\\
		0&\Omega_-[\psi_1,\psi_2]
	\end{matrix}\right)\left(\begin{matrix}
		\psi_1\\
		\psi_2
	\end{matrix}\right)=0,
\end{align}
where $\beta=\beta_1+\beta_2$, $\varphi=2\theta_{\bf B}$, and $\Omega_{\pm}[\psi_1,\psi_2]\equiv 2\lambda_1(|\psi_1|^2+|\psi_2|^2)\pm 2\lambda_2(|\psi_1|^2-|\psi_2|^2)$.
Without loss of generality, we assume the initial state to be $\psi_1$ before applying the magnetic field, i.e., a $d+id$ superconductor. After applying the magnetic field, $\psi_2$ is also mixed into the ground state. However, this admixture is small when the field is weak.
On the contrary, when the applied field is sufficiently strong
\begin{align}
	B\gtrsim\frac{\hbar}{2e}\frac{1}{d}
	\sqrt{\frac{3\alpha}{2(\gamma-\beta})}\equiv B_c,
\end{align} 
the two order parameters $\psi_1$ and $\psi_2$ are driven to be the same in magnitude, and the ground state is no longer a chiral one. We call the complete loss of chirality a phase transition since the symmetry of the phase is completely changed. We note that $B\rightarrow 0$ when $\alpha\rightarrow 0$ near $T_c$. With the ansatz $\psi_2=\psi_1 e^{i\varphi}$, we find 
\begin{align}
	|\psi_{1,2}|^2=\frac{1}{4\lambda_1}\left[-\alpha+(\gamma-\beta)\left(\frac{2e}{\hbar}\right)^2{\bf A}^2\right],
\end{align}
which is suppressed by orbital effects when $\beta>\gamma$, as is the case in MATBG (see numerical results below).

The order parameters with equal contribution of $\psi_{1,2}$ can be generally represented by 
\begin{align}
\tilde{\psi}_{1,2}=(\psi_1\pm e^{-i\varphi}\psi_2)/\sqrt{2}.
\label{new_OP}
\end{align}
 When decomposing the condensate Boson field $\Phi=\tilde{\psi}_1\tilde{\xi}_1+\tilde{\psi}_2\tilde{\xi}_2$ [Eq.~(\ref{decomposition})], we can define new basis $\tilde{\xi}_{1,2}=(\xi_1\pm e^{i\varphi}\xi_2)/\sqrt{2}$ for $\tilde{\psi}_{1,2}$, respectively, which lie at the equator of the Bloch sphere [Fig.~\ref{fig:Bloch}(b)]. Since $\psi_2=\psi_1e^{i\varphi}$, the new state driven by the in-plane magnetic field is exactly $\tilde{\psi}_1$.  
 We can thereby obtain its Lagrangian by substituting the transformation Eq.~(\ref{new_OP}) into Eq.~(\ref{effective_L}), yielding  
 the linearized free energy of $\tilde{\psi}_1$
 \begin{align}
 	F=\int d{\bf r}\tilde{\psi}_1^*({\bf r}){\cal H}(\hat{\bf r},t)\tilde{\psi}_1({\bf r}).
 	\label{effective_L}
 \end{align}
 Here we define 
 \[{\cal H}(\hat{\bf r},t)=\alpha-\sum_{\mu\nu}\tilde{c}_{\mu\nu}\left(\partial_{\mu}-\frac{2e}{i\hbar }{\bf A}_{\mu}\right)\left(\partial_{\nu}-\frac{2e}{i\hbar }{\bf A}_{\nu}\right),
 \]
where the components of the stiffness are
\begin{align}
	\nonumber
	\tilde{c}_{xx}&=\beta+\gamma\cos(2\theta_{\bf B}),\\
	\nonumber
	\tilde{c}_{yy}&=\beta-\gamma\cos(2\theta_{\bf B}),\\
	\tilde{c}_{xy}&=\tilde{c}_{yx}=\gamma\sin(2\theta_{\bf B}).
	\label{coefficients_c}
\end{align} 
Intriguingly, these components are tunable by the direction of in-plane magnetic field, and are anisotropic along the $\hat{\bf x}$- and $\hat{\bf y}$-directions, viz., correspond to emerging nematicity that breaks the three-fold rotation symmetry in the magnetic-field-driven phase.

\textit{Nematic paraconductivity.}---The field-driven phase that is the ground state below $T_c$ shows a nematic transport response as addressed below.  Slightly \textit{above} the superconducting transition temperature, the conductivity, called paraconductivity, is mainly contributed by the superconductor order parameters since their fluctuation under thermal noise can carry a supercurrent \cite{Tinkham,Nagaosa}. The measurement of DC resistivity around $T_c$, as performed in MATBG \cite{Pablo_exp}, can thus reflect the fluctuation of the superconducting order parameters and provide information about the pairing symmetry of the superconducting state. To calculate the response to an electric field, the vector potential in ${\cal H}(\hat{\bf r},t)$ is increased by ${\bf A}_E=-{\bf E}t$,
which is the contribution of the applied electric field \cite{Nagaosa}. Then by the free energy (\ref{effective_L}), the time-dependent GL equation, augmented by thermal noise, is written as \cite{Tinkham,Nagaosa}
\begin{align}
	\Gamma\partial_t\tilde{\psi}_1({\bf r},t)=-{\cal H}(\hat{\bf r},t)\tilde{\psi}_1({\bf r},t)+f({\bf r},t),
	\label{TDGL_noise}
\end{align}
where $\Gamma$ is the damping rate for the superconducting order parameter $\tilde{\psi}_1({\bf r},t)$ and $f({\bf r},t)$ represents thermal noise. We assume that the thermal noise is white with correlation relation  
$\langle f^*({\bf r},t)f({\bf r}',t')\rangle=2\Gamma k_BT\delta({\bf r}-{\bf r}')\delta(t-t')$ \cite{Tinkham,Nagaosa}.
Via a Fourier transformation, the electric current reads
\begin{align}
	\nonumber
	{\bf J}&\equiv -\frac{\delta F}{{\cal S}\delta {\bf A}_E}\\
	&=-\frac{2k_BT}{\Gamma {\cal S}}\sum_{\bf q}\Lambda({\bf q})\int_{-\infty}^0du\exp\left(-\frac{2}{\Gamma}\int_{u}^0dt{\cal H}({\bf q},t)\right),
\end{align}
where ${\cal S}$ is the sample area and $\Lambda({\bf q})\equiv{\partial {\cal H}({\bf q},t)}/{\partial {\bf A}_E}|_{{\bf A}_E\rightarrow 0}$. From this the paraconductivity is determined in linear response to be 
\begin{align}
	\sigma_{ij}=k_BT\frac{e^2\Gamma}{2\pi\hbar^2\alpha}\frac{\tilde{c}_{ij}}{\sqrt{\beta^2-\gamma^2}},
\end{align}
which is a tensor that exhibits a Hall response. This emerging Hall effect is unique to the magnetic-field-driven phase since it is absent in the chiral superconducting phase without magnetic field, which follows an isotropic paraconductivity $\sigma^c_{ij}= k_BTe^2\Gamma/({2\pi\hbar^2\alpha})\delta_{ij}$.
In particular, when the electric field is applied along the $\hat{\bf x}$-direction in the coordinate system defined in Fig.~\ref{fig:Bloch}(a), the induced current along the $\hat{\bf x}$-direction is modulated as 
$\tilde{c}_{xx}$, and there is a Hall response with the current along the $\hat{\bf y}$-direction being modulated as $\tilde{c}_{yx}$. They are both nematic with a two-fold anisotropy [Eq.~(\ref{coefficients_c})].

\begin{figure}[ht]
	\begin{center}
		{\includegraphics[width=4.1cm,trim=0.25cm -0.7cm 0.5cm 0cm]{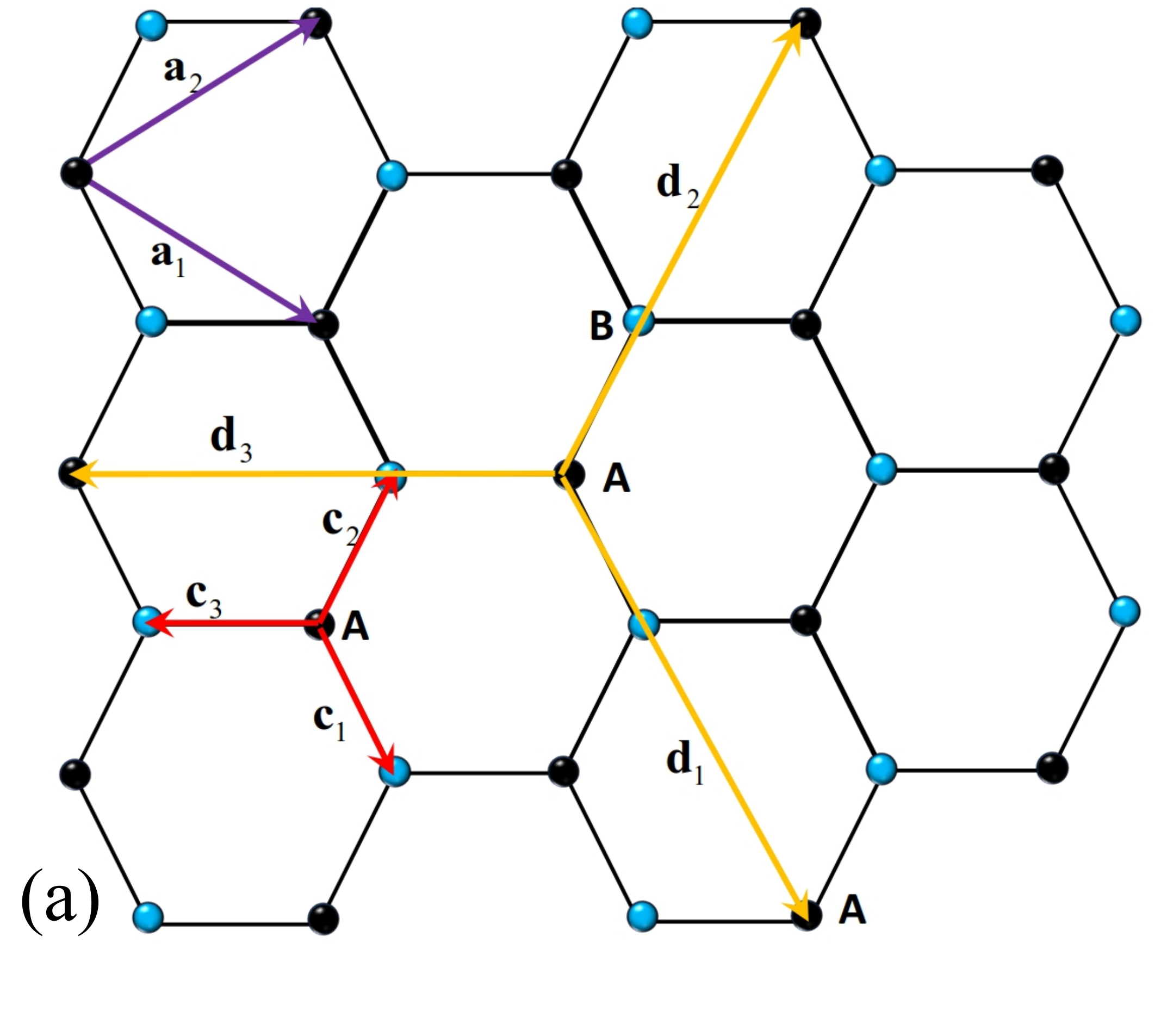}}
		{\includegraphics[width=4.35cm,trim=0.0cm 0cm 0.45cm 0cm]{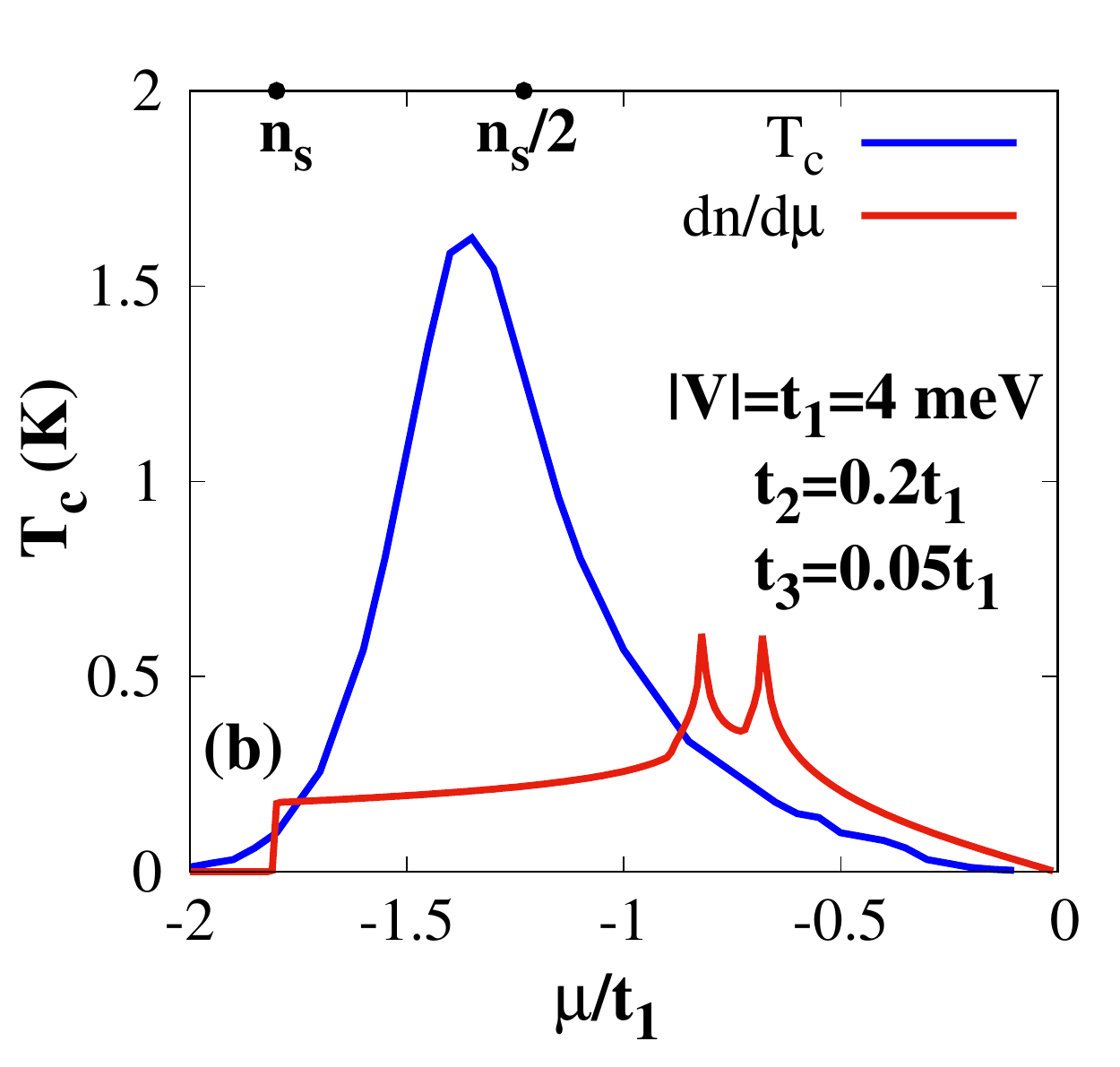}}
		{\includegraphics[width=4.25cm,trim=0.15cm 0cm 0.5cm 0cm]{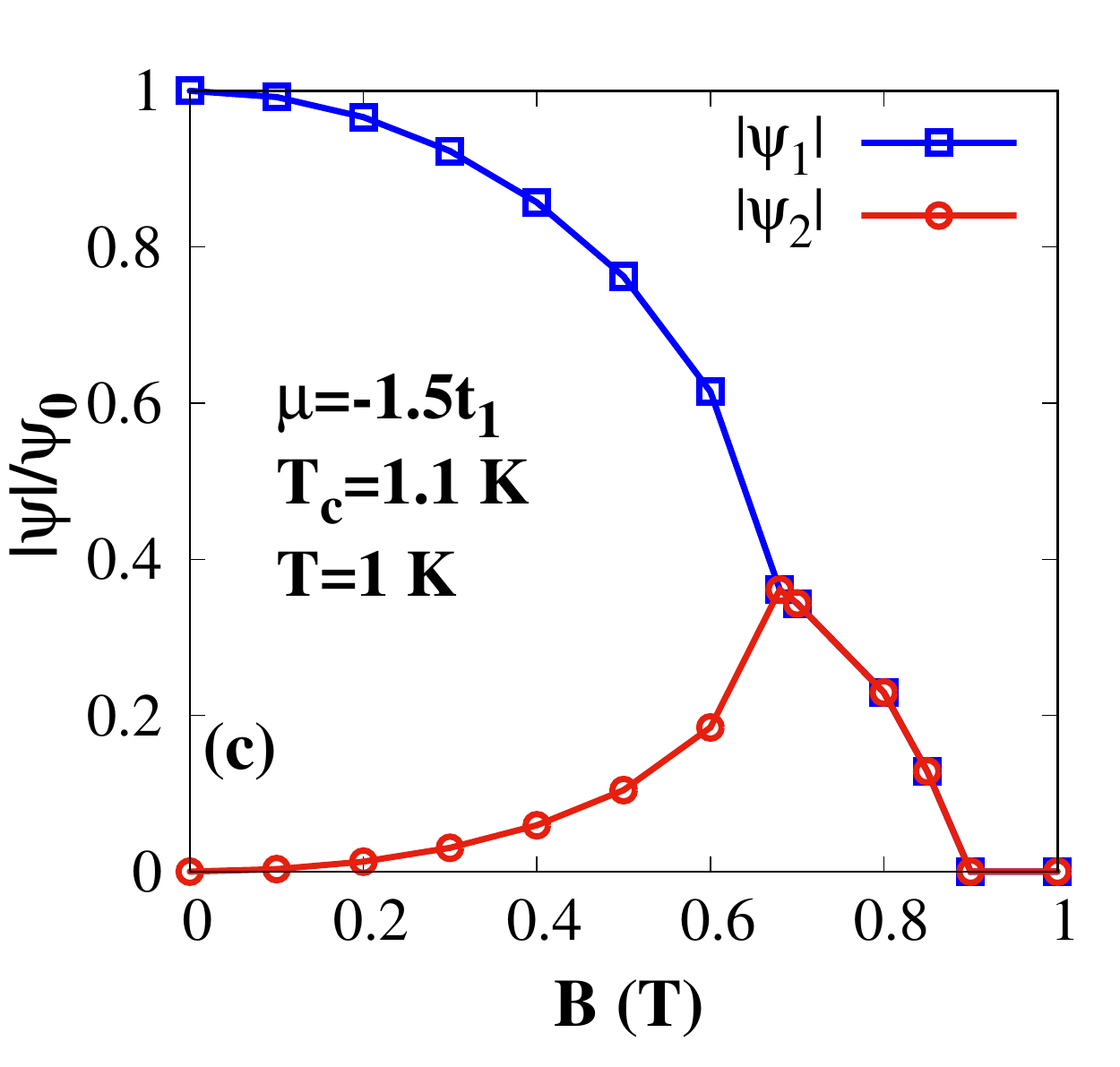}}
		{\includegraphics[width=4.2cm,trim=0.12cm 0cm 0.5cm 0cm]{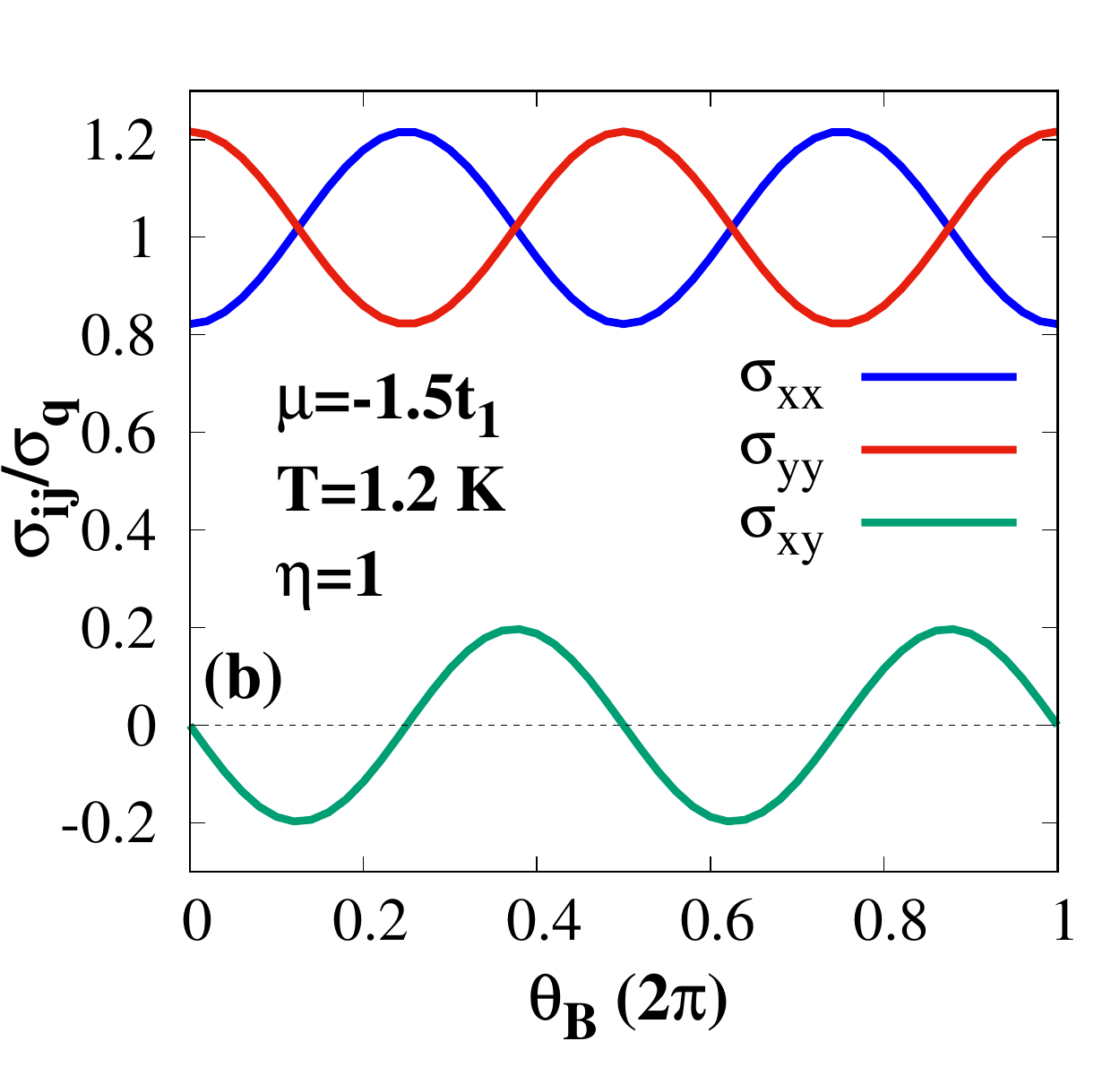}}
		\caption{Microscopic model for calculation of GL parameters and paraconductivity. (a), the Moir\'e honeycomb lattice, where ${\bf a}_{1,2}$ are two Bravais lattice vectors, and ${\bf c}_{1,2,3}$ and ${\bf d}_{1,2,3}$ represent the nearest and fifth neighboring bonding vectors. (b), calculated critical superconducting temperature and compressibility. The superconducting dome with largest $T_c\approx 1.5$~K recovers the typical features in the experiment \cite{Pablo_exp}. (c), the phase transition induced by in-plane magnetic field when $T\lesssim T_c$, in which $\psi_0=0.063$~meV and a phase transition at in-plane magnetic field $B\sim 0.6$~T is predicted. (d), the dependence of the paraconductivity on the magnetic-field direction.}  
		\label{fig:dome}
	\end{center}
\end{figure}

\textit{Parameter estimation.}---To be specific for the estimation of GL parameters and paraconductivity, here we consider the tight-binding model on the honeycomb lattice with two $p$-orbitals $\{p_x,p_y\}$ on every site proposed by Yuan and Fu \cite{Yuan_Fu,Yuan_Fu2,Kang_Jian_PRX}, as shown in Fig.~\ref{fig:dome}(a). The superlattice has a point group of $D_3$.
 Note, however, that our general conclusions rely on the above symmetry analysis only and are thus applicable beyond this specific microscopic model. The chiral basis is denoted by  $(\hat{a}_{{\bf k}\pm,\sigma},\hat{b}_{{\bf k}\pm,\sigma})^T=(\hat{a}_{{\bf k}x,\sigma}\pm \hat{a}_{{\bf k}y,\sigma},\hat{b}_{{\bf k}x,\sigma}\pm \hat{b}_{{\bf k}y,\sigma})^T/\sqrt{2}$ on the ``A" and ``B" sites with $\sigma=\{\uparrow,\downarrow\}=\{+,-\}$ being the electron spin, the Hamiltonian in momentum space is divided into subspaces, given by
\begin{align}
\nonumber
{h}_{\pm,\sigma}({\bf k})&=-\mu+t_2(g_{\bf k}+g_{-{\bf k}})
\pm it_3(g_{-{\bf k}}-g_{\bf k})\\
&+\frac{\sigma}{2}g\mu_B B+\left(\begin{matrix}
0&t_1f_{\bf k}\\
t_1f_{-{\bf k}}&0,
\end{matrix}\right).
\label{kinetic}
\end{align}
Here, $\mu$ is the chemical potential; $t_1$ and $\{t_2,t_3\}$ are the hopping parameters between nearest and fifth neighboring sites that are connected by ${\bf c}_{\mu=\{1,2,3\}}$ and ${\bf d}_{\mu=\{1,2,3\}}$ [Fig.~\ref{fig:dome}(a)], respectively \cite{Yuan_Fu}; $f({\bf k})=\sum_{\mu={1,2,3}}e^{i{\bf k}\cdot{\bf c}_{\mu}}$ and $g({\bf k})=\sum_{\mu={1,2,3}}e^{i{\bf k}\cdot{\bf d}_{\mu}}$; and $g\mu_B B$ is the Zeeman splitting via the in-plane magnetic field. We disregard the Zeeman effect since its suppression of the superconductivity is weak when $B\rightarrow 0$,  while the orbital effect is still relevant when $T\rightarrow T_c$.
We use the interaction Hamiltonian that allows to stabilize $d$-wave superconductivity \cite{Martin,RMP,switching,graphene_NP,Honerkamp},
\begin{align}
\nonumber
\hat{H}_{\rm int}
&\simeq\sum_{\alpha=\pm}\sum_{{\bf k}{\bf k}'}V({\bf k}-{\bf k}')\left(\hat{a}_{{\bf k}\alpha,\uparrow,}^{\dagger}\hat{b}^{\dagger}_{-{\bf k}\alpha,\downarrow}-\hat{a}_{{\bf k}\alpha,\downarrow}^{\dagger}\hat{b}^{\dagger}_{-{\bf k}\alpha,\uparrow}\right)\\
&\times\left(\hat{b}_{-{\bf k}'\alpha,\downarrow}\hat{a}_{{\bf k}'\alpha,\uparrow}-\hat{b}_{-{\bf k}'\alpha,\uparrow}\hat{a}_{{\bf k}'\alpha,\downarrow}\right),
\label{interaction}
\end{align}
where the pairing potential 
\[
V({\bf k}-{\bf k}')=\frac{V}{N}\sum_{\mu=\{1,2,3\}}e^{i({\bf k}-{\bf k}')\cdot{\bf c}_{\mu}}.
\]
Here $N$ is the number of honeycomb lattice sites, and the coupling constant $V<0$.

For this specific model, the condensate boson field  contains three components $\Vec{\Phi}=(\phi_1,\phi_2,\phi_3)^T$ when we consider the pairing between three nearest neighbors on the honeycomb lattice, with which the effective Lagrangian  \cite{path_integral} (see Supplemental Material \cite{supplement} for details), 
\begin{align}
\nonumber
&L_{\rm eff}[\overline{\phi},\phi]=\sum_{\mu\mu'=\{1,2,3\}}\int d{\bf r}\overline{\phi}_{\mu}({\bf r},\tau){\cal M}_{\mu\mu'}\phi_{\mu'}({\bf r},\tau)\\
\nonumber
&+\sum_{\mu\mu'}\sum_{\delta\gamma=x,y}{\cal T}_{\delta\gamma}^{\mu\mu'}\int d{\bf r}\partial_{\delta}\phi_{\mu}^*({\bf r},\tau)\partial_{\gamma}\phi_{\mu'}({\bf r},\tau)+O(\phi^4),
\end{align}
where ${\cal M}_{\mu\mu'}$ determines the superconducting transition, and ${\cal T}_{\delta\gamma}^{\mu\mu'}$  controls the spatial fluctuations. In this model, the basis functions $\xi_{1}=-\left(e^{-i\frac{2\pi}{3}},1,e^{ i\frac{2\pi}{3}}\right)^T/\sqrt{3}$ and  $\xi_{2}=\left(1,e^{-i\frac{2\pi}{3}},e^{-i\frac{4\pi}{3}}\right)^T/\sqrt{3}$, with the phase difference $\pm 2\pi/3$ rooted in the angle difference $\pm 120^{\circ}$ between the nearest bonding vectors. With the basis function, we find the mass $a_i\equiv\xi_i^{\dagger}\pmb{\cal M}\xi_i$ and stiffness $c^{ij}_{\delta\gamma}\equiv\xi_i^{\dagger}\pmb{\cal T}_{\delta\gamma}\xi_j$ of the order parameters, and calculate the GL parameters via relations, e.g., $a_1=a_2=\alpha$, $c_{\delta\gamma}^{11}=c_{\delta\gamma}^{22}=\beta\delta_{\delta\gamma}$ and $c_{xx}^{21}=-c_{yy}^{21}=-ic_{xy}^{21}=\gamma$ \cite{supplement}.

We first estimate the magnitude of the in-plane magnetic field to realize the phase transition to the nematic superconducting state in MATBG. With parameters $|V|=t_1=4$~meV, $t_2=0.2t_1$, $t_3=0.05t_1$ and $|{\bf c}_{\mu}|=14/\sqrt{3}~{\rm nm}$ for the Moir\'e honeycomb lattice \cite{NS1,NS2}, the critical temperature $T_c$ of chiral $d$-wave superconductivity is calculated by solving $\alpha(T_c)=0$. $T_c$ is shown in Fig.~\ref{fig:dome}(b) and exhibits a dome with a peak at hole doping $n_h>n_s/2$ where $n_s=4/\Omega$ characterizes doping four holes in one Moir\'e unit cell of area $\Omega$ \cite{Pablo_exp,NS1}.
We note that this $T_c$ is well below the BKT transition within our mean-field framework \cite{supplement}. 
This peak is not at the Van Hove points of the band that are characterized by two peaks in compressibility $dn/d\mu$ with our parameters. With a typical hole doping at $\mu=-1.5t_1$, we estimate $T_c\approx 1.1$~K,  $\alpha=-10^{-4}/|{\bf c}_{\mu}|^2~{\rm meV}^{-1}\cdot{\rm m}^{-2}$, $\beta=2.6~{\rm meV}^{-1}$, $\gamma=-0.5~{\rm meV}^{-1}$, and $\lambda_1=-3\lambda_2=0.02/|{\bf c}_{\mu}|^2~{\rm meV}^{-3}\cdot{\rm m}^{-2}$ at temperature $T=1$~K \cite{supplement,Yuan_Fu}. With the thickness $d\approx0.8$~nm of TBG estimated by roughly twice the single-layer one $\sim 0.34$~nm \cite{graphene_RMP}, the two order parameters $\psi_{1,2}$ become close in magnitude when the applied magnetic field $B\gtrsim 0.6$~T, as shown in Fig.~\ref{fig:dome}(c), which is close to the typical value $B_c\approx 0.5$~T found experimentally \cite{Pablo_exp}.

We then estimate the paraconductivity by choosing $\Gamma=\eta\alpha\hbar/(k_BT_c)$ with a factor $\eta$ of order $1$ \cite{Gorkov,Kopnin}, leading to a universal paraconductivity $\sigma_{ij}=\eta \sigma_q\tilde{c}_{ij}/\sqrt{\beta^2-\gamma^2}$ with $\sigma_q=e^2/(2\pi\hbar)=3.87\times 10^{-5}$~S being the conductance quantum. With the parameters at $\mu=-1.5t_1$ and $T=1.2$~K, we plot $\sigma_{ij}/\sigma_q$ in Fig.~\ref{fig:dome}(d) with $B\gtrsim 0.6$~T and $\eta$ taken to be 1. The conductivities oscillate with respect to the magnetic field with period of $\pi$, thus exhibiting a two-fold anisotropy. The amplitude of the oscillation is determined by $|\gamma|$ that may depend on the behind microscopic mechanism. The direction of the in-plane magnetic field tunes the sign of $\sigma_{xy}$ and hence the direction of the Hall current, which also could provide an intriguing functionality for future applications.

\textit{Discussion.}---We have demonstrated  nematic paraconductivity that  emerges in two-dimensional chiral superconductors under an in-plane magnetic field. This effect is particularly instructive for the chiral $d$-wave superconducting state of the honeycomb lattice in that the driven phase shows two-fold anisotropy that breaks the three-fold one of the lattice. Furthermore, the 
magnetic-field--driven nematic phase shows a Hall effect in a non-ferromagnetic system. The underlying mechanism relies on the hybridization of chiral order parameters by an in-plane vector potential with shifted the nodes of the gap in the driven phase, which could be directly tracked by STM \cite{STM_RMP}. Our work has direct implications for the pairing symmetry of superconductivity in MATBG.  Since our purely symmetry-based mechanism applies in the general context of two-dimensional superconducting order parameters it might be relevant to experimental observations in other materials, such e.g., few-layer NbSe$_2$ reported recently \cite{Hamill,footnote}, as well.

\vskip0.25cm 
\begin{acknowledgments}
We thank Rafael Fernandes and Liang Fu for useful discussions. 
TY and MAS acknowledge financial support by Deutsche Forschungsgemeinschaft through the Emmy Noether program (SE 2558/2). DMK acknowledges support by the Deutsche
Forschungsgemeinschaft (DFG, German Research Foundation) via RTG 1995, within the Priority Program SPP 2244 ``2DMP'' and Germany’s Excellence Strategy - Cluster of Excellence Matter and Light for Quantum Computing (ML4Q) EXC 2004/1 - 390534769. AR  acknowledges support from the European Research Council (ERC- 2015-AdG-694097), UPV/EHU Grupos Consolidados (IT1249-19) and  the Cluster of Excellence `CUI: Advanced Imaging of Matter' of the Deutsche Forschungsgemeinschaft (DFG) - EXC 2056 - project ID 390715994. The Flatiron Institute is a division of the Simons Foundation. We acknowledge support from the Max Planck-New York City Center for Non-Equilibrium Quantum Phenomena.

\end{acknowledgments}


%

\end{document}